**Title:** Design and Analysis of Extended Depth of Focus Metalenses for Achromatic Computational Imaging

**Author List:**

Luocheng Huang[1,+], James Whitehead[1,+], Shane Colburn[1], & Arka Majumdar[1,2,*]

[1]Department of Electrical and Computer Engineering, University of Washington, Seattle, Washington 98195, USA.

[2]Department of Physics, University of Washington, Seattle, Washington 98195, USA.

[*]Correspondence to: arka@uw.edu

[+]These authors contributed equally

**Abstract:**

Metasurface optics have demonstrated vast potential for implementing traditional optical components in an ultra-compact and lightweight form factor. Metasurface lenses, also called metalenses, however, suffer from severe chromatic aberrations, posing serious limitations on their practical use. Existing approaches for circumventing such aberrations via dispersion engineering are limited to small apertures and often entails multiple scatterers per unit cell with small feature sizes. Here, we present an alternative technique to mitigate chromatic aberration and demonstrate high-quality, full-color imaging using extended depth of focus (EDOF) metalenses and computational reconstruction. Previous EDOF metalenses relied on cubic phase masks that induced asymmetric artifacts in images, whereas here we demonstrate the use of symmetric phase masks that can improve subsequent image quality, including logarithmic-aspherical, and shifted axicon masks. Our work will inspire further development in achromatic metalenses beyond dispersion engineering and open new research avenues on hybrid optical-digital metasurface systems.

**TOC Graphic:**

**Introduction:**

Over the last decade, image sensors have undergone dramatic miniaturization, thanks to advances in optical packaging and semiconductor-based photodetector technology. Even further size reduction, however, is required for emerging areas of machine vision, autonomous transportation, and augmented reality visors[1–3]. Such miniaturization using traditional refractive optics is difficult, as the optical elements themselves occupy a significant volume. An attractive solution to reduce the overall volume of these imaging systems is to use subwavelength diffractive optics, also known as metasurfaces. Metasurfaces are two-dimensional optical elements consisting of quasi-periodic subwavelength resonators that are capable of abruptly introducing phase shifts to a wavefront, enabling ultrathin optics and lenses[4–11]. Unfortunately, these metasurfaces exhibit severe chromatic aberrations, a long-standing problem in diffractive optics. For a metalens, the focal length is inversely proportional to the wavelength, originating primarily from the fixed positions of phase-wrapping discontinuities as the wavelength changes[12]. This shift in focal length with a change in wavelength causes chromatic aberrations that blur the image. Recent works have attempted to mitigate this chromatic aberration through dispersion engineering[13–18], which employs scatterers that compensate for the chromatic phase dispersion. The phase delay at each scatter has a wavelength dependence that is effectively corrected using higher order terms in the Taylor expansion of the phase with frequency. Dispersion-engineered metasurfaces, however, are limited to a small aperture for a fixed numerical aperture (NA)[13]; a larger aperture would require a higher maximum phase dispersion, which requires the optical resonators to have higher quality factors. This would imply that the scatterers would need ever higher aspect ratios for increased aperture size, exceeding current nanofabrication capabilities.

Another technique for mitigating chromatic aberrations is to employ freeform metasurfaces and computational imaging[19–23] a paradigm that does not have the aforementioned scaling challenges although this process entails additional energy consumption and latency due to the need of computational reconstruction. With advancements in and the availability of fast and efficient computing, however, this can be done swiftly and with minimal energy consumption. Thus, computational imaging in conjunction with freeform metasurfaces is a promising avenue for mitigating metalens aberrations. Recently, full-color imaging in the visible wavelength regime was demonstrated using an extended depth of focus (EDOF) metasurface and post-processing deconvolution[20]. Here, a rectangularly separable cubic phase mask (CPM)[24] was added to the standard hyperboloidal metalens phase, generating a non-rotationally symmetric extended focal spot. The extended nature of the focal spots at different wavelengths is sufficient to compensate for the chromatic shift in the focal length. The EDOF property of the CPM enables the imaging system to capture useful spatial frequency information of the colored image so that computational reconstruction is possible.[20] The CPM is limited, however, in that it produces an asymmetric PSF that makes imaging sensitive to the orientation of the element, often manifesting as asymmetric artifacts even after deconvolution. Additionally, the CPM produces a lateral shift of the PSF with a change in wavelength, which can contribute to distortions in imaging. One potential solution to this limitation is to utilize a rotationally symmetric point spread function. Although there are previous works doing so based on extending the depth of focus of refractive optics, using log-asphere[25] and axicon-based lenses[26], they have not been used for correcting the strong chromatic aberrations encountered in a metasurface platform.

In this paper, we extend the family of EDOF metasurfaces beyond a simple CPM. We design and fabricate four different types of EDOF metasurface lenses operating in the visible regime, including both rotationally symmetric and asymmetric phase profiles. We characterize the MTF for all of these lenses and demonstrate full-color imaging. A comparative analysis of all these lenses, is also provided, in terms of optical bandwidth and imaging performance. All our EDOF metasurfaces demonstrate at least an order of magnitude larger optical bandwidth. Full-color imaging in the visible range is achieved using all the EDOF lenses, outperforming the traditional metalens in terms of chromatic aberrations.

**Results:**

*Design and Fabrication:*

An imaging system behaves as a linear system that maps the incoming light from a scene to the sensor. This mapping function of the optical element is modelled by a convolution with the element's point spread function (PSF). For imaging under incoherent light, the system can be considered to provide a linear mapping from the input intensity to the output intensity, which is captured on the sensor array. By scanning a point source throughout the object volume and measuring the resultant intensity across the image volume, a 3D intensity impulse response, or PSF, can be measured. This method can fully characterize the imaging function of the system. Assuming the PSF is shift-invariant, we can treat the PSF as a kernel that convolves with the input to produce an image on the sensor plane. For an ideal lens, the PSF resembles a point, which enables capturing an exact replica of the scene. When the PSF deviates from a point, the captured image becomes blurry; however, with a PSF known a priori, an in-focus image can be retrieved via post-capture deconvolution if sufficient spatial frequency information is captured by the sensor. This information retrievability can be expressed in terms of the modulation transfer function (MTF) of the optical element, which is given by a slice of the modulus of the Fourier transform of the PSF.

A broad MTF, one that does not drop to zero rapidly, signifies that a wide range of spatial frequencies is captured at the sensor plane, corresponding to PSF with a small spot size. On the other hand, a narrow MTF, one that decays rapidly, captures only a limited range of spatial frequency content, which precludes the possibility of computational reconstruction due to the zeros in the spatial frequency spectrum. A conventional, in-focus metalens exhibits a broad MTF when imaging with narrowband light, resulting in high-quality images. When imaging with a different wavelength at the same sensor plane, however, the spatial bandwidth of the MTF drastically decreases and the collected spatial frequencies from the scene are attenuated or eliminated. As some of the spatial frequencies are not collected, this results in an uncorrectable blur. With EDOF lenses, we can realize a similar PSF at the sensor plane for a broad and continuous range of wavelengths[27]. Moreover, the resulting MTF can capture a wider range of spatial frequencies, which is key to generating full-color images at high resolution.

Here, we designed four different EDOF lenses, namely EDOF cubic, shifted axicon, log-asphere, and SQUBIC lenses. All, except for the cubic, are axially symmetric. We fix the aperture for each of these metasurfaces at 200 μm and select a nominal focal length of 200 μm, making the numerical aperture (NA) close to 0.45 for all designs. We emphasize that the small aperture of the metalens is due to the prohibitive cost and time for fabricating large aperture lenses based on electron-beam lithography, and not due to scaling limitations as encountered with dispersion-engineered metalenses. The cubic metalens utilizes a focusing phase mask combined with a cubic term to produce a MTF insensitive to wavelength.[20] Thus, the phase mask for an EDOF cubic lens is:

$$\phi(x, y) = \frac{2\pi}{\lambda} \left( \sqrt{x^2 + y^2 + f^2} - f \right) + \frac{\alpha}{R^3} (x^3 + y^3) \tag{1}$$

where $\lambda$ denotes the operational wavelength, $x$ and $y$ are the coordinates in plane, $f$ is the nominal focal length, $\alpha$ represents the strength of the cubic term, and $R$ represents the radius of the phase mask. We chose $\lambda = 550$ nm, $f = 200$ μm, $R = 100$ μm, and $\alpha = 55\pi$ for the design of the cubic metasurface. To compare the performance of the EDOF metasurfaces against the standard singlet metalens, we include a design with $\alpha = 0$ that imparts no cubic term to the wavefront. The log-asphere phase mask was inspired by a prior work[25] that divided the phase mask into annular zones with continuously varying focal lengths. The central annular zone has a focal length of $s_1$ and the outermost annular zone has a focal length of $s_2$. This design effectively extends the focal length from $s_1$ to $s_2$. The log-asphere phase mask is governed by the relation:

$$\phi(r) = \int_0^r \frac{r}{\left\{ r^2 + \left[ s_1 + (s_2 - s_1)\left(\frac{r}{R}\right)^n \right]^2 \right\}^{1/2}} \tag{2}$$

where $r = \sqrt{x^2 + y^2}$ and $R$ is the aperture radius of the phase mask. The parameter $n$ changes the intensity distribution over the optical axis. For the log-asphere phase mask, we set $n = 2$, making the intensity distribution uniform across the line of foci. The parameters we chose for the log-asphere metasurface design are $\lambda = 550$ nm, $s_1 = 80$ μm, $s_2 = 300$ μm, and $R = 100$ μm. Similar to the log-asphere phase mask, the shifted axicon phase mask takes the same form of the Eq. (2). In this case, we set $n = 1$, $\lambda = 550$ nm, $s_1 = 80$ μm, $s_2 = 300$ μm. The axial intensity distribution of the shifted axicon resembles that of a diffractive axicon lens[28], hence the nomenclature. The SQUBIC metasurface was first demonstrated by Patway to have EDOF properties[29]. The phase mask directs a collimated light beam into a line of discrete foci, achieving an EDOF, and is given by the equation:

$$\phi(x,y) = 2\pi A \left[ \frac{\sqrt{1-\left(\left(\frac{x}{R}\right)^2+\left(\frac{y}{R}\right)^2\right)\sin^2\alpha} - 1}{1-\cos\alpha} + \frac{1}{2} \right]^3 \qquad (3)$$

where $\alpha = \sin^{-1}(NA)$, given that $NA$ is the nominal numerical aperture of the optics. Here, $A$ is a design parameter that determines the strength of the phase mask[29]. In our design, we set $A = 50$ and NA = 0.45. Fig. 1A shows the wrapped phase distributions of all the metasurfaces.

These EDOF lenses are then implemented using cylindrical $Si_3N_4$ nanopillars[11,30] to ensure polarization insensitivity. These nanopillars are arranged on a square lattice. By varying the diameters of these nanoposts, the transmission coefficient imparted on incident light is modified as the coupling to and amongst different supported modes by the nanoposts changes, resulting in different phase shifts. Rigorous coupled-wave analysis (RCWA) is used to construct a library consisting of the diameters of the nanoposts and the corresponding phase shift and amplitude.

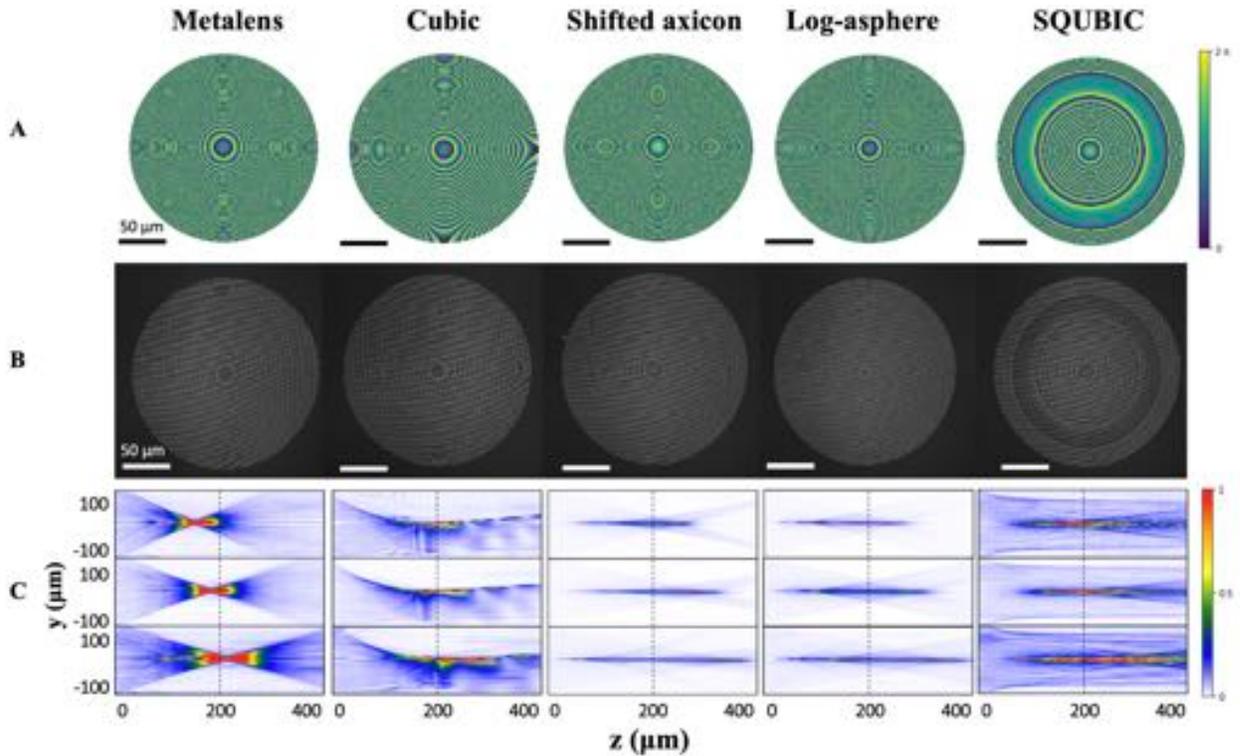

*Figure 1: EDOF Metasurface Design and Measurements: (A) The phase masks of an ordinary metalens and four different EDOF metasurfaces. (B) Scanning electron micrographs of the fabricated metasurfaces. (C) We experimentally measured the intensity along the optical axis where from top to the bottom panel represents illumination by 625, 530, and 455 nm wavelengths.*

## *Characterization and Imaging:*

The fabricated metasurfaces (one ordinary metalens and four EDOF metasurfaces) are then characterized experimentally. A fiber-coupled LED light source (Thorlabs M625F2, M530F2, M455F1) is used for illumination of the metasurfaces. A custom microscope mounted on a computer-controlled translation stage is used to take snapshots of the imaging plane. The focal length of the metalens is found to be 231 μm with the green LED. This deviation from the designed focal length (200 μm) could be due to the difference in the designed wavelength (550 nm) and the experimental light source wavelength (530 nm). The imperfect fabrication could have also contributed to this deviation. The nominal focal length is thereafter set to 231 μm for all five metasurfaces to accommodate this shift. To verify the extended depth of focus, PSF slices are measured along the optical axis (z-axis) spanning 0 to 400 μm from metasurface to construct a 3D PSF, shown in the Fig. 1C. The PSF measurement on the x-y plane is shown in Fig. 2, under different wavelength illumination. We note that for a standard metalens, the PSFs are very different for different colors, whereas, for the EDOF lenses the variation is significantly reduced. To quantitatively understand this behavior, we calculate the MTF of all the metasurfaces at the nominal focal plane. As can be seen in Fig. 2, the MTF of an ordinary metalens well preserves spatial frequency information for green light (when focused at the sensor plane) but fails to preserve high frequency components for blue or red. The cubic metasurface presents a broad frequency preservation, exhibiting a higher cutoff frequency than the standard metalens. Similarly, the other EDOF metasurfaces demonstrate an oscillatory behavior: while there are some zeros, for several high spatial frequency components we have non-zero MTF over a wider range compared to the standard metalens. This also plays an important role, as when an image is captured under white light illumination, all the wavelengths are captured at the sensor array, and having a similar MTF enables computational reconstruction, as discussed below.

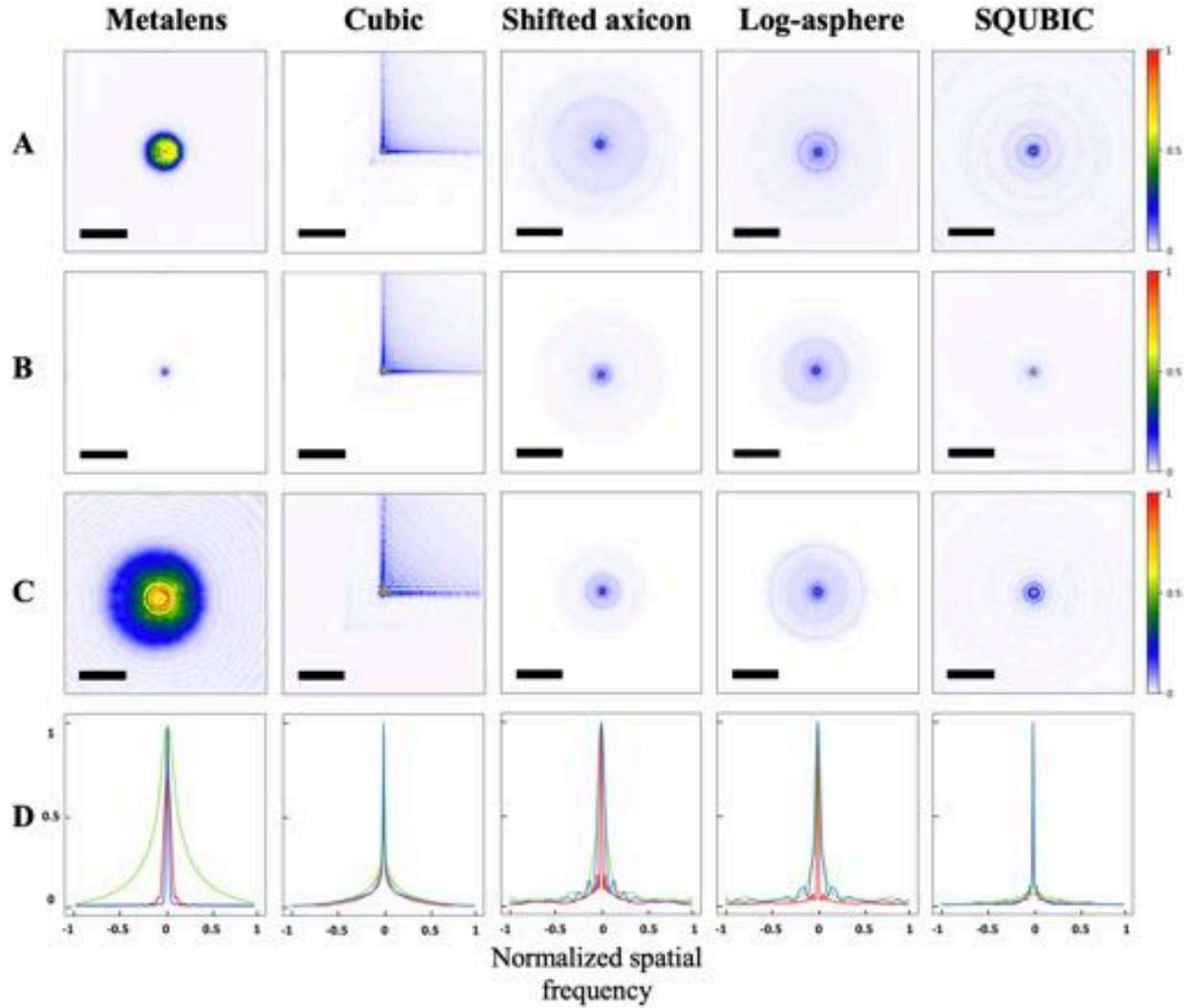

*Figure. 2. **Characterization of the metasurfaces**. The PSFs of the singlet metasurfaces were measured under 455 nm blue (A), 530 nm green (B), and 625 nm red (C). The corresponding MTFs are displayed with red line from its PSF measured under red light, green lines under green light, and blue lines under blue light. (D) The scale bar signifies 25 μm.*

After characterization of the PSFs and MTFs, we captured colored images using all the metasurfaces. The same setup is used for imaging with an addition of an OLED monitor (SmallHD

5.5" Focus OLED HDMI Monitor). The OLED monitor is placed ~15 cm away from the metasurface, which displays images shown on Fig. 3. The images are then captured using a camera (Allied Vision GT1930C). These raw images are then computationally reconstructed. Our EDOF imaging system presents a problem in the form of $O = Kx + n$, in which $O$ is the observed image, $K$ is the PSF or blur kernel, $x$ is the latent image, and $n$ is noise that corrupts the captured data. Although several methods exist to restore the image $x$, given $K$ and $O$, we chose to use Wiener deconvolution[29] due to its effectiveness and simplicity. The normal metalens exhibits such a high level of chromatic aberration that we were unable to restore the raw images with any reasonably good results.

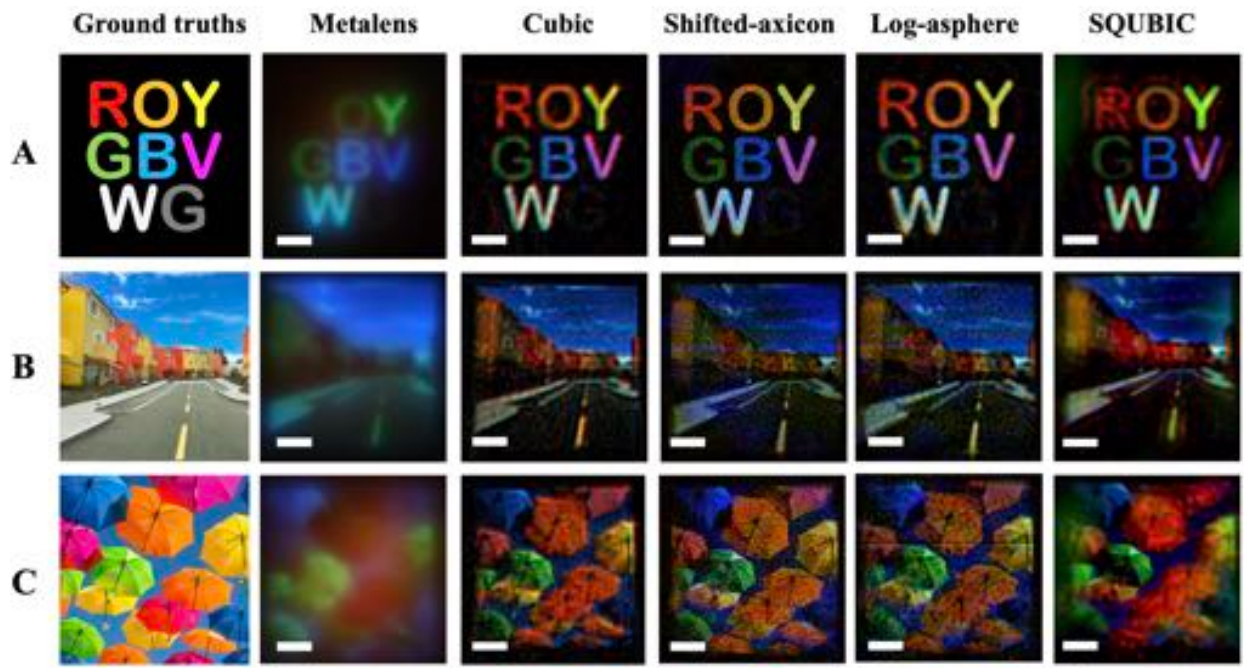

*Figure. 3.* **Imaging performance**. *Restored images taken from an OLED display of colored letters in ROYGBVWG (A), a colorful neighborhood (B), and vibrant umbrellas against the sky. (C) The scale bar signifies 20 μm. Note that the metalens images are raw and unrestored.*

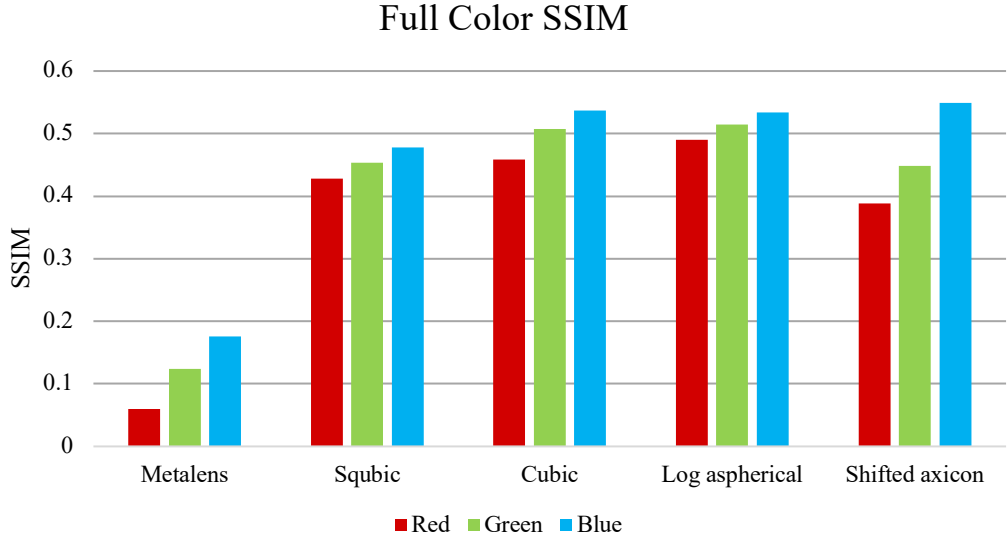

*Figure. 4.* **Full Color SSIM**. *The restored captures are scaled, rotated, and translated to align with the ground truth then SSIM is calculated for each color channel for the metasurface.*

To obtain an objective imaging quality comparison across the five metasurfaces, structural similarity (SSIM) tests are carried out using the "ROYGBVWG" image shown on Fig. 3 (A). The SSIM has an upward trend from red to blue which appears correlated with the similar upward trend in focusing efficiency (see supplement Table S1). Since the SSIM algorithm is sensitive to alignment, the captured images are aligned to the ground truth (GT) through a similarity transformation[32] (rotation, translation, and scaling) using an image registration algorithm (see more details in the supplement). The SSIMs are then calculated between the GT and the output images for the red, green, and blue channel respectively. The values of the SSIM range from 0 to 1, 1 being a perfect match. As shown in Fig. 4, the imaging system with the singlet metalens exhibits the lowest SSIM score compared to the rest of the metasurfaces. Even though the SSIM score is at most 0.55, the scores on the devices with EDOF properties are significantly higher than that of the standard metalens. The metalens obtained poor imaging quality throughout all colors, although designed to work for green. The low SSIM of the metalens for green is due to the fact that the OLED display used has a nominal wavelength (511 nm) different from that of the light source (531 nm) used to measure the PSF for all the colors. The OLED monitor also has a much larger bandwidth (~35 nm for green) compared to the simulated bandwidth for the metalens, leading to a blurry image even for the intended wavelength. (see the supplementary materials for the spectra of the light sources).

Table. 1. **Imaging bandwidth**, defined as the bandwidth at one half of the maximum PSF similarity coefficient.

|  | Metalens | Log-asphere | Shifted Axicon | Cubic | SQUBIC |
|---|---|---|---|---|---|
| Bandwidth (nm) | 15.2 | 233.3 | 233.3 | 112.1 | 157.6 |
| Center Wavelength (nm) | 543.9 | 553.0 | 547.0 | 540.9 | 530.3 |

As discussed before, it is desirable to have an invariant PSF across the full visible wavelength range to achieve full-color imaging. To capture the color invariance of the PSFs on the metasurfaces, we perform PSF correlation calculations with simulated data that estimate the spectral bandwidth. The correlation coefficient is calculated as the inner product between the baseline PSF and the test PSF (see the supplementary materials). Table 1 shows the bandwidth (measured as the optical bandwidth where the correlation values falls to half of the maximum[33]) and the central wavelength of all the metasurfaces. Compared with the EDOF metasurfaces, the standard metalens has a much narrower spectral bandwidth, as is also evident from the MTF and SSIM measurements.

**Discussion:**

The SSIM results agree with the poor imaging quality exhibited by the metalens in the visible regime, caused by its strong chromatic aberrations. In contrast, the EDOF metasurfaces demonstrate an impressive ability to maintain a highly invariant PSF across a large spectral bandwidth that makes them an excellent platform for computational imaging. The imaging results as well as the SSIM calculations also indicate that EDOF metasurfaces significantly outperform the standard metalens in full-color imaging. The log-asphere and shifted axicon designs both demonstrate the highest optical bandwidth for imaging among the tested EDOF designs. We note that recently a fundamental limit on the achievable optical bandwidth is reported given a thickness and numerical aperture of the lens.[34] For our parameters the fundamental limit becomes around ~100nm (see supplementary materials), which is smaller than all the bandwidth demonstrated using EDOF lenses. While it is indeed possible to increase the optical bandwidth at the expense of image quality, it is difficult to estimate the Strehl ratio for our EDOF lenses as the final image is obtained only after deconvolution.

Our algorithm takes ~2 seconds to complete the task of image restoration, which includes loading of the raw image as well as the PSF image, and exporting to file, all using a single CPU. This process can, however, be significantly accelerated by utilizing multiple CPUs, or a dedicated field-programmable gate array. Preloading the PSF into memory will also speed up the computational reconstruction process. This work also uses Wiener deconvolution due to its speed, however, more sophisticated deconvolution algorithms can be used to improve the image quality at the expense of time[35,36], the most costly of which could be applied offline.

This work demonstrates the viability of four different metasurface designs for full-color imaging with minimal chromatic aberrations. We design and fabricate all of these EDOF metasurfaces, which when combined with deconvolution, correct for chromatic aberrations caused by the inherent diffractive nature of the metasurface. Imaging and analyses are carried out on our devices that compare the performance between the traditional metalens and the four different EDOF metasurfaces. Although previous works have demonstrated achromatic imaging with dispersion-engineered metasurfaces, our imaging system does not rely on any input polarization state and is generalizable to larger aperture applications. A larger aperture will enable higher signal-to-noise-ratio and faster shutter speed, which are crucial to any commercial application. Moreover, dispersion-engineered metasurfaces require high aspect ratio scatterers for large apertures and also have multiple meta-atoms per unit cell, making the fabrication more challenging. In our approach the aspect ratio of the scatterers is relatively small, and thus our metasurfaces have more relaxed fabrication constraints.

Compared with the singlet metalens, our EDOF imaging system exhibits an exceptional ability to maintain imaging fidelity under broadband illumination. Our imaging platform combines the form factor of ultrathin diffractive metasurfaces and the flexibility of computational imaging, making our platform an attractive solution for novel imaging applications. The CMOS compatibility of our silicon nitride platform, combined with a high NA of ~0.40, makes this approach ideal for miniaturized microscopy, smartphone cameras, and endoscopy. Moreover, it is possible to increase the aperture of our EDOF hybrid system while maintaining the same NA and imaging characteristics. This aperture scalability opens avenues to applications such as planar cameras and even satellite imaging.

**Methods:**

The designed metasurfaces are fabricated on the same sample. First, a double side polished fused silica wafer is cleaned with acetone and IPA. Plasma-enhanced chemical vapor deposition is used to deposit 623 nm of Silicon nitride (via SPTS). A layer of 200 nm of ZEP-520A is then spun on the wafer and 8 nm of Au/Pd is sputtered for charge dissipation. The design pattern is written using electron-beam lithography (JEOL JBX6300FS at 100 kV). Then the Au/Pd layer is removed by immersing in Gold Etchant Type TFA (Transene) and the chip is developed in amyl acetate. Next,

50 nm of aluminum is evaporated onto the developed pattern and then lifted off, leaving a patterned aluminum etch mask. The silicon nitride layer is etched through its full thickness in an inductively-coupled plasma etcher using a fluorine chemistry (Oxford Plasmalab 100). Finally, the aluminum is removed, producing the metasurfaces. Fig. 1B shows the scanning electron micrograph (SEM) of all the metasurfaces.

**Acknowledgement:**

This work is supported by Samsung-GRO, UW reality lab, Google, Huawei and Facebook, and NSF-1825308. Part of this work was conducted at the Washington Nanofabrication Facility / Molecular Analysis Facility, a National Nanotechnology Coordinated Infrastructure (NNCI) site at the University of Washington, which is supported in part by funds from the National Science Foundation (awards NNCI-1542101, 1337840 and 0335765), the National Institutes of Health, the Molecular Engineering & Sciences Institute, the Clean Energy Institute, the Washington Research Foundation, the M. J. Murdock Charitable Trust, Altatech, ClassOne Technology, GCE Market, Google and SPTS.

**Supporting Information Available:**

Image Restoration

SSIM Evaluations

PSF Correlation Coefficient Calculations

Focusing Efficiency

Optical Setup

Light Source Spectra

**Title:** Design and Analysis of Extended Depth of Focus Metalenses for Achromatic Computational Imaging: Supplementary Materials

**Author List:**

Luocheng Huang[1,+], James Whitehead[1,+], Shane Colburn[1], & Arka Majumdar[1,2,*]

[1]Department of Electrical and Computer Engineering, University of Washington, Seattle, Washington 98195, USA.

[2]Department of Physics, University of Washington, Seattle, Washington 98195, USA.

[*]Correspondence to: arka@uw.edu

[+]These authors contributed equally

1. **Image Restoration**

To restore the captured raw images given the point spread function (PSF), a 3×3 median filter is first applied to the raw image to remove hot pixels. Then, Wiener deconvolution is applied to the image. We use the Wiener-Hunt deconvolution algorithm [1], which takes in the raw image and the PSF. Before is it passed into the algorithm, the PSF is normalized by dividing by the sum of all PSF pixels. The actual implementation uses the *skimage.restoration.wiener* module from the *Scikit-image* library [2]. The Wiener filter estimates the desired image $\hat{x}$ as follows:

$$\hat{x} = \mathcal{F}^{-1}(|\Lambda_H|^2 + \lambda|\Lambda_D|^2)\Lambda_H^\dagger \mathcal{F} y,$$

where $\Lambda_H$ is the optical transfer function, $\Lambda_D$ is a Laplacian filter that penalizes high frequencies, and $\lambda$ is the parameter that tunes the regularization. The $\lambda$ used for the deconvolutions is $1 \times 10^{-4}$.

2. **SSIM Evaluations**

SSIM calculations are sensitive to the subject's translation, scaling, and rotation, which are difficult to eliminate in an experimental setup. Hence, image registration is performed using the MATLAB image registration module for optimal alignment. An intensity-based image registration method is utilized to transform the captured image to align with the ground truth using a similarity transformation. This operation is performed for all three color channels simultaneously for each capture.

3. **PSF Correlation Coefficient Calculations**

First, a wavelength-sensitive phase mask is generated consisting of transmission coefficients obtained through rigorous coupled-wave analysis, using the Stanford S4 package [3]. Then, the wavefront is propagated to the image plane using the angular spectrum method. The correlation coefficient is calculated as the inner product between the baseline PSF and the test PSF at the test wavelength, normalized to the maximum coefficient. The plots are presented in Fig. S1.

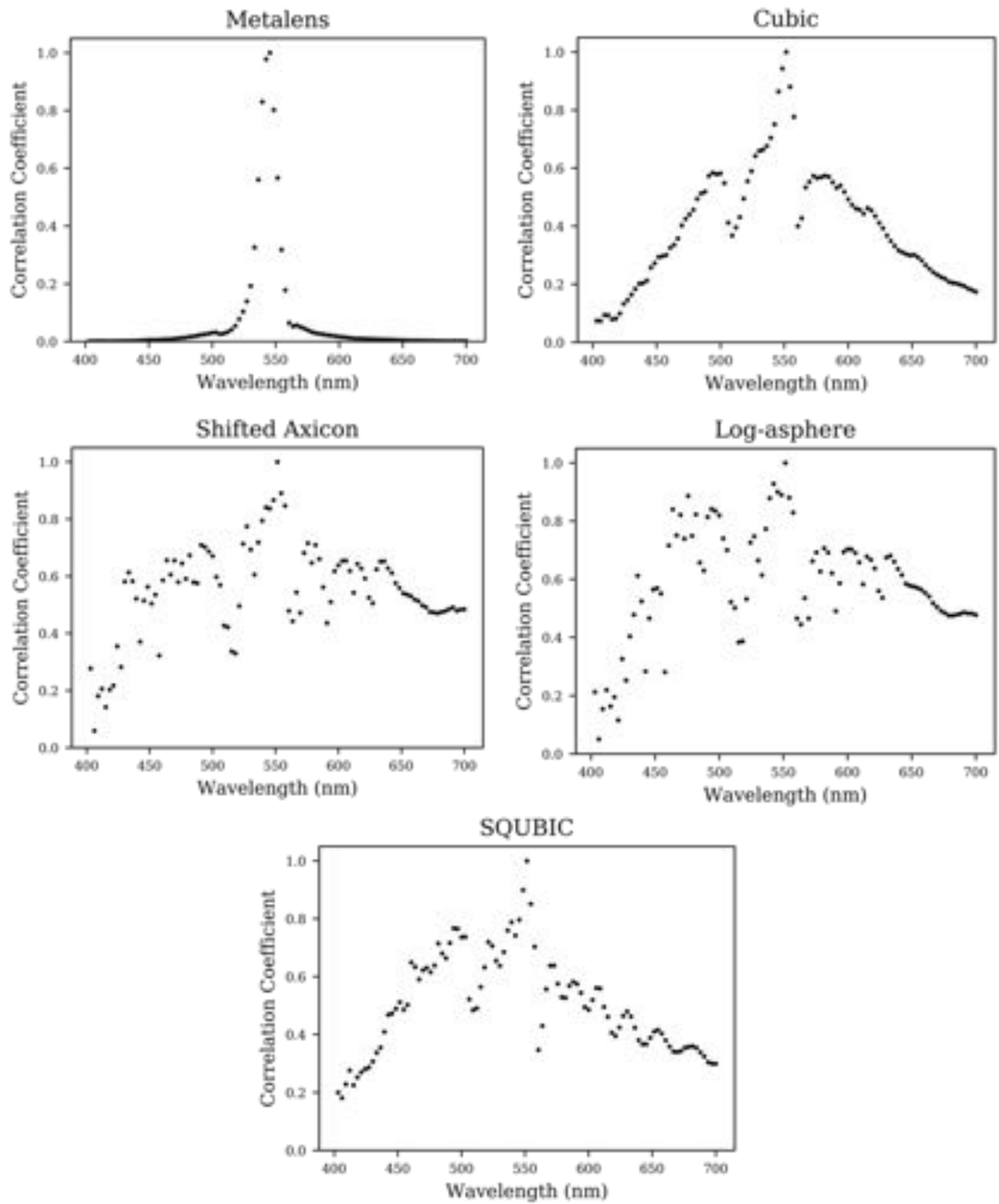

**Fig. S1.** Correlation coefficient plots. The Y-axis is the correlation coefficient, X-axis is the wavelength.

4. **Focusing Efficiency**

We define the focusing efficiency as the power ratio of the light at the focal plane to the light transmitted through the metalens. To calculate the efficiency, we first capture an image at the focal plane, such that no pixel is overexposed (i.e., without any saturation). Then, using the same setting, we capture an image at the plane of the metasurface. We then define a circular mask where the metasurface spans and sum all the pixels in this region to determine the total transmitted power. Finally, the ratio of the focal plane power and the transmitted power yields the focusing efficiency. The measured efficiencies are presented on Table 1. We note that we have defined this efficiency in such a fashion as several of the EDOF lenses do not produce a single defined intensity lobe by design, making efficiency calculation methods based on a single lobe impractical for comparison.

**Table S1.** Focusing efficiencies of the metasurfaces

|       | Metalens | Cubic | Log-asphere | Shifted axicon | SQUBIC |
|-------|----------|-------|-------------|----------------|--------|
| Blue  | 67.7%    | 86.2% | 93.0%       | 96.3%          | 91.1%  |
| Green | 45.5%    | 85.4% | 85.7%       | 87.5%          | 89.9%  |
| Red   | 93.0%    | 88.8% | 70.7%       | 66.9%          | 96.8%  |

5. **Optical Setup**

The optical setup for capturing PSFs is illustrated in Fig. S2 and the imaging setup is illustrated in Fig. S3.

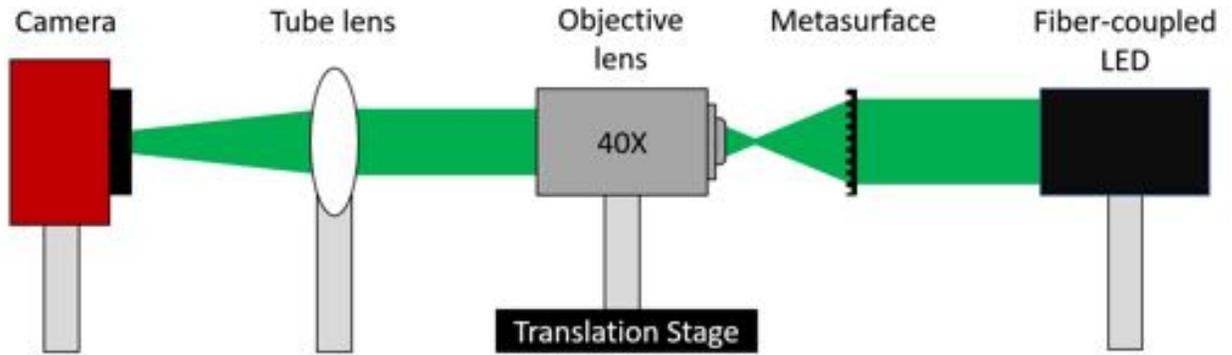

**Fig. S2.** The experimental setup for PSF measurements. A collimated beam from a fiber-coupled LED (Thorlabs M625F2, M530F2, and M455F1) is directed at the metasurface. The collimated beam is then focused by a metasurface lens, producing a PSF at the working distance of the 40x objective (Nikon Plan Fluor, NA = 0.75, WD = 0.66 mm). Finally, a tube lens (Thorlabs ITL200) projects and magnifies the image onto a camera (Allied Vision Prosilica GT 1930C) that saves the raw captures for deconvolution.

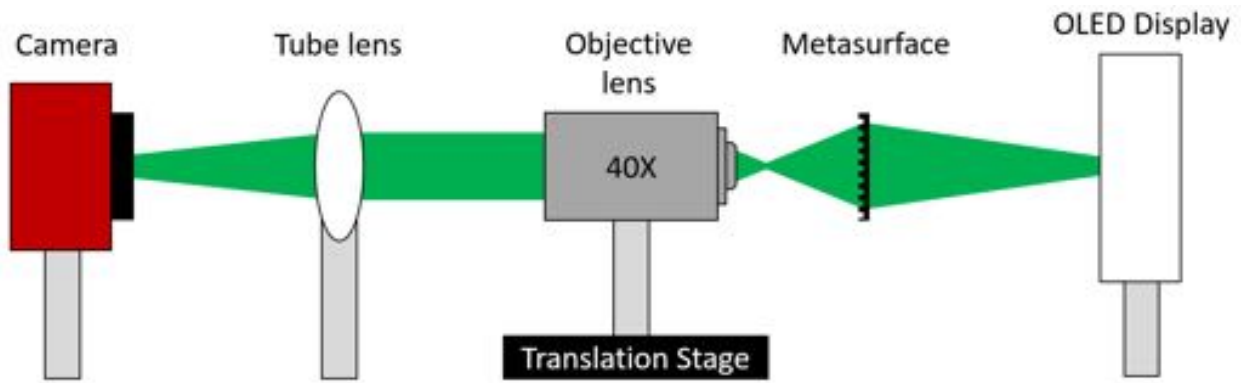

**Fig. S3.** The experimental setup for imaging. An OLED monitor (SmallHD 5.5" FOCUS OLED Monitor) displays a pattern. The incident OLED signal is then collected and focused by a metasurface lens, producing an image at the working distance of a 40x objective (Nikon Plan Fluor, NA = 0.75, WD = 0.66 mm). Finally, a tube lens (Thorlabs ITL200) projects and magnifies the image onto a camera (Allied Vision Prosilica GT 1930C) that saves the raw captures for deconvolution.

6. **Light Source Spectra**

We have recorded the spectra of the OLED monitor (SmallHD 5.5" FOCUS OLED Monitor) as well as the fiber-coupled LEDs. The individual spectra are recorded when displaying a single color at a time using (IsoPlane SCT320). The intensity for each color is normalized to one and are presented in Fig. S4. The fiber-coupled LEDs (Thorlabs M625F2, M530F2, and M455F1) are also measured and displayed in Fig. S5.

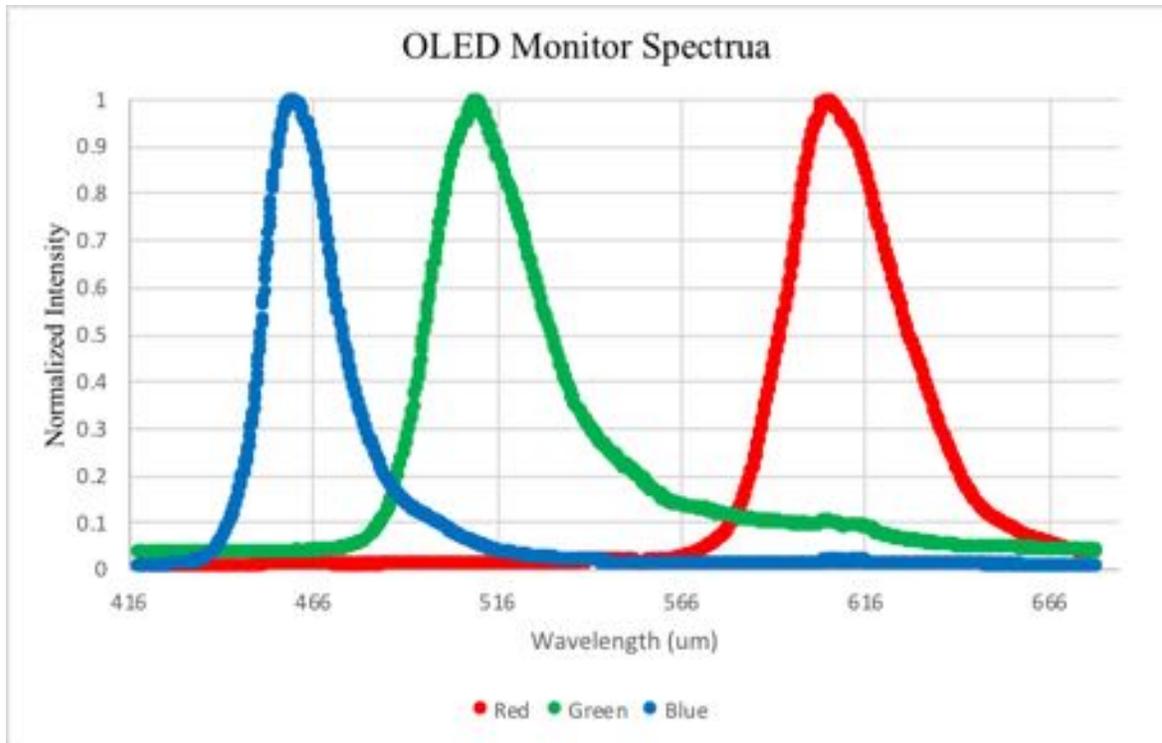

**Fig. S4.** Measured spectra of individual color channels on SmallHD 5.5" FOCUS OLED Monitor. (IsoPlane SCT320) is used for the measurement.

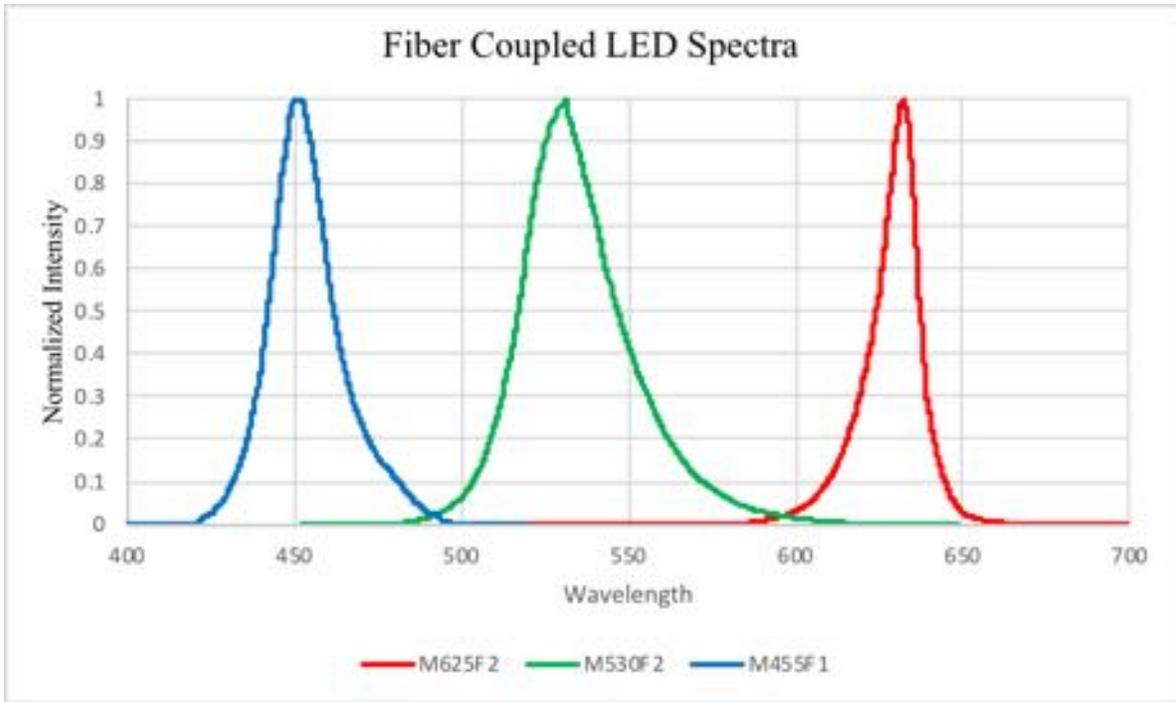

**Fig. S5.** Measured LED spectra of the fiber coupled LEDs: M625F2, M530F2, and M455F1 for red, green, and blue PSF measurements.

7. **Theoretical Bandwidth Calculations**

The theoretical bandwidth of each lens is calculated to better understand its bandwidth performance using eq. (S2). [4]

$$\Delta \omega \leq \omega_c \frac{L \Delta n}{f} \Theta\left(\frac{NA}{n_b}\right), \tag{S1}$$

Where $\omega_c$ is the central frequency, L is the height of the pillars, $f$ is the focal length, $\Delta n$ is the refractive index difference between the background and the pillars, $n_b$ is the background refractive index, and NA is the numerical aperture of the lens. The equation (S2) is used to substitute into equation (S1), leading to the theoretical bandwidths $\Delta \omega$ for each metasurface, listed on table S2.

$$\Theta\left(\frac{NA}{n_b}\right) = \frac{\sqrt{1 - \left(\frac{NA}{n_b}\right)^2}}{1 - \sqrt{1 - \left(\frac{NA}{n_b}\right)^2}} \tag{S2}$$

**Table S2.** Theoretical Bandwidths of EDOF Metasurfaces

|  | Metalens | Cubic | Log-asphere | Shifted axicon | SQUBIC |
|---|---|---|---|---|---|
| Bandwidth (μm) | 104.4 | 106.2 | 105.0 | 103.8 | 101.8 |